# MLSA4Rec: Mamba Combined with Low-Rank Decomposed Self-Attention for Sequential Recommendation

Jinzhao Su, Zhenhua Huang*

School of Computer Science, South China Normal University, Guangzhou, China.

## Abstract

In applications such as e-commerce, online education, and streaming services, sequential recommendation systems play a critical role. Despite the excellent performance of self-attention-based sequential recommendation models in capturing dependencies between items in user interaction history, their quadratic complexity and lack of structural bias limit their applicability. Recently, some works have replaced the self-attention module in sequential recommenders with Mamba, which has linear complexity and structural bias. However, these works have not noted the complementarity between the two approaches. To address this issue, this paper proposes a new hybrid recommendation framework, **M**amba combined with **L**ow-Rank decomposed **S**elf-**A**ttention for Sequential **Rec**ommendation (MLSA4Rec), whose complexity is linear with respect to the length of the user's historical interaction sequence. Specifically, MLSA4Rec designs an efficient Mamba-LSA interaction module. This module introduces a low-rank decomposed self-attention (LSA) module with linear complexity and injects structural bias into it through Mamba. The LSA module analyzes user preferences from a different perspective and dynamically guides Mamba to focus on important information in user historical interactions through a gated information transmission mechanism. Finally, MLSA4Rec combines user preference information refined by the Mamba and LSA modules to accurately predict the user's next possible interaction. To our knowledge, this is the first study to combine Mamba and self-attention in sequential recommendation systems. Experimental results show that MLSA4Rec outperforms existing self-attention and Mamba-based sequential recommendation models in recommendation accuracy on three real-world datasets, demonstrating the great potential of Mamba and self-attention working together.

# 1 Introduction

Sequential recommender systems play a vital role in applications such as e-commerce, online education, and streaming services [1,2,3]. By analyzing users' historical interaction data, these systems can predict the next possible interaction, thereby providing personalized recommendations. This not only improves user satisfaction but also significantly enhances the platform's revenue.

Sequential recommender systems have evolved from traditional methods to deep learning models. Early sequential recommendation methods mainly relied on collaborative filtering and matrix factorization techniques [4,5]. Although these methods are simple, they perform poorly when dealing with long sequences and sparse data. With the rise of deep learning, researchers began exploring the use of neural networks to improve the performance of sequential recommender systems. Methods based on convolutional neural networks (CNNs) [6,7] can effectively capture local features through local connections and weight sharing mechanisms. This performs well in handling short sequences or tasks with strong local relevance. However, CNNs have limitations in capturing sequential order information, making it difficult to fully reflect user preferences. Methods based on recurrent neural networks (RNNs), such as GRU4Rec [8] and NARM [9], can effectively capture temporal dependencies in sequences through the recursive structure of RNNs and their variants. However, RNNs may encounter gradient vanishing or exploding problems when processing sequences [10], and their computational efficiency is relatively low. Methods based on self-attention mechanisms, such as SASRec [11] and BERT4Rec [12], calculate dependency scores between each pair of items in the sequence through self-attention mechanisms [13], performing well in capturing dependencies between items in user historical interactions. Compared to CNNs and RNNs, self-attention mechanisms can capture dependencies between all items in the user interaction sequence without relying on fixed window sizes or recursive structures. Additionally, self-attention mechanisms can process sequence data in parallel, improving computational efficiency. However, their quadratic complexity makes these methods computationally inefficient and costly in handling long sequences [14]. Moreover, these methods lack structural bias, as self-attention mechanisms do not assume any structural priors in the input. This makes self-attention-based methods prone to overfitting noise, thereby affecting their generalization ability [15]. One study [13] has also experimentally confirmed that the accuracy of the vanilla self-attention-based Transformer [22] in certain tasks is much lower than that of state-of-the-art methods with structural bias.

To overcome these issues, recent studies [16,17] have begun to focus on replacing the self-attention module in sequential recommenders with Mamba, which has linear complexity and structural bias. Mamba [18] is a type of state space model (SSM) whose unique selection mechanism allows it to filter noise and retain relevant information based on the input sequence, standing out among many SSMs. Although Mamba performs well in long-sequence modeling, it has limitations in capturing local information, restricting its performance in recommendation tasks. Moreover, these methods merely regard Mamba as a replacement for self-attention mechanisms, ignoring the possibility of their complementarity. Some works [19,20,21] in other research fields have noticed this and have proven the potential of hybrid frameworks combining both through extensive experiments. However, these works still use the vanilla self-attention mechanism with quadratic complexity, which inevitably drags down Mamba's efficiency. Inspired by this, this paper proposes a novel hybrid sequence recommendation framework with linear complexity, called MLSA4Rec.

This framework achieves accurate and efficient sequence recommendations by combining low-rank decomposition self-attention (LSA) [14] and Mamba. The core of MLSA4Rec is the Mamba-LSA interaction module, where Mamba injects structural bias into LSA, providing it with global and sequential information. Subsequently, LSA, through a gated information transfer mechanism, offers Mamba critical information from a different perspective, such as the user's potential interests or local information that Mamba might overlook. Finally, linear transformations and activation functions are used to integrate the outputs resulting from the interactions between the two, and the Mamba normalization layer refines the hybrid outputs containing multi-source knowledge for the next item prediction.

In summary, the main contributions of this paper are as follows:

- To the best of our knowledge, this is the first study to explore the potential of Mamba and self-attention mechanisms working together in sequential recommendation, and even in the entire recommendation system.
- We propose a novel hybrid framework MLSA4Rec, which achieves complementary advantages of Mamba and self-attention mechanisms, effectively capturing comprehensive and multi-perspective dependencies in user interaction items.
- We conduct extensive experiments on multiple real-world datasets to verify the superior performance of MLSA4Rec in sequential recommendation tasks.

## 2. Preliminaries

### 2.1 Definition of Sequential Recommendation Task

The sequential recommendation task aims to predict the next possible interaction of a user based on the user's historical interaction sequence. Specifically, given a user set $U = \{u_1, u_2, ..., u_{|U|}\}$ and an item set $V = \{v_1, v_2, ..., v_{|V|}\}$, for a given user $u$ in $U$ with historical interaction sequence $S_u = [v_1, v_2, ..., v_{n\_u}]$, the sequential recommendation model needs to predict the next interaction item $v_{n\_u+1}$ for user $u$.

### 2.2 State Space Models

SSMs are inspired by the classical state space model [23], having linear or near-linear computational complexity with respect to sequence length, and have been widely applied in fields such as video, audio, and time series [24]. The foundation of SSMs is the linear ordinary differential equation, which can transform a one-dimensional continuous input signal $x(t) \in \mathbb{R}$ into a one-dimensional continuous output signal $y(t) \in \mathbb{R}$:

$$\begin{aligned} h'(t) &= \mathbf{A}h(t) + \mathbf{B}x(t) \\ y(t) &= \mathbf{C}h(t) \end{aligned} \quad (1)$$

where $h(t) \in \mathbb{R}^N$ is the hidden state, and $\mathbf{A} \in \mathbb{R}^{N \times N}, \mathbf{B} \in \mathbb{R}^{N \times 1}, \mathbf{C} \in \mathbb{R}^{1 \times N}$ are learnable parameters.

To handle discrete sequences $x = [x_1, x_2, ..., x_L] \in \mathbb{R}^L$, some works, such as S4 [25], choose to discretize the parameters in the above equation, resulting in:

$$\begin{aligned} h_t &= \overline{\mathbf{A}} h_{t-1} + \overline{\mathbf{B}} x_t \\ y_t &= \mathbf{C} h_t \end{aligned} \quad (2)$$

where discrete parameters $\overline{\mathbf{A}}, \overline{\mathbf{B}}$ are calculated from continuous parameters $\mathbf{A}, \mathbf{B}$ through zero-

order hold formula:

$$\overline{\mathbf{A}} = \exp(\Delta \mathbf{A}),$$
$$\overline{\mathbf{B}} = (\Delta \mathbf{A})^{-1}(\exp(\Delta \mathbf{A}) - \mathbf{I}) \cdot \Delta \mathbf{B}. \qquad (3)$$

where the step size $\Delta$ can be regarded as the resolution of continuous input $x(t)$. After completing the discretization, SSMs can be used in a linear recursive manner that is equivalent to convolution for efficient computation [26].

**2.3 Mamba**

The linear time-invariant nature of SSMs leads to parameters in the equations being unchanged based on input or time, limiting their context-based reasoning ability. To solve this issue, Mamba introduces an input-dependent selection mechanism, adding the ability to selectively propagate or forget information along the sequence. Specifically, $\mathbf{B}$, $\mathbf{C}$, and $\Delta$ are constructed based on the input sequence $\mathbf{X} \in \mathbb{R}^{L \times D}$ through linear projections:

$$\mathbf{B}, \mathbf{C}, \Delta = \text{Linear}(\mathbf{X}) \qquad (4)$$

where $\mathbf{B} \in \mathbb{R}^{L \times N}$, $\mathbf{C} \in \mathbb{R}^{L \times N}$, $\Delta \in \mathbb{R}^{L \times D}$, and $L$ is the sequence length, $D$ is the number of channels. Equations (1), (2), (3), and (4) together form the selective SSM.

Mamba is built on this selective SSM, consisting of multiple Mamba blocks, which are alternately arranged with standard normalization layers and residual connections. Through selective SSM and efficient hardware-aware algorithm design, Mamba achieves powerful modeling capabilities while maintaining scalability, providing a new option for sequential modeling tasks.

**2.4 Low-Rank Decomposed Self-Attention Mechanism**

Low-rank decomposed self-attention mechanism (LSA) can be used to generate context-aware item representations. This module projects items into $P$ latent interests and fuses contextual information through the interaction between items and each latent interest. This avoids direct interaction between items, effectively reducing the computational complexity of the self-attention mechanism from $O(L^2)$ to $O(PL)$ [14]. LSA mainly includes two operations: item-to-interest aggregation and interest-to-item interaction.

**2.4.1 *Item-to-Interest Aggregation.*** LSA assumes that the interactive items of most users can be categorized into $P$ latent interests and uses a learnable projection function $f : \mathbb{R}^{L \times D} \to \mathbb{R}^{P \times D}$ to project items into these latent interests. Specifically, given the item embedding matrix $\mathbf{H} \in \mathbb{R}^{L \times D}$, LSA first calculates the relevance distribution $\mathbf{Z}$ from items to interests and aggregates the item embedding matrix through $\mathbf{Z}$, resulting in the interest representation matrix $\widetilde{\mathbf{H}}$:

$$\widetilde{\mathbf{H}} = f(\mathbf{H}) = \mathbf{Z}^\top \cdot \mathbf{H} = \left(\text{softmax}\left(\mathbf{H} \cdot \mathbf{\Theta}^\top\right)\right)^\top \cdot \mathbf{H} \qquad (5)$$

where $\mathbf{\Theta} \in \mathbb{R}^{P \times D}$ is a learnable parameter matrix. This process converts the item embedding matrix $\mathbf{H}$ into a low-rank interest representation $\widetilde{\mathbf{H}} \in \mathbb{R}^{P \times D}$, effectively reducing the size of the attention matrix and improving the computational efficiency of attention.

**2.4.2 *Item-to-Interest Interaction.*** The input embedding sequence $\mathbf{X}$ is transformed into three matrices $\mathbf{Q}, \mathbf{K}, \mathbf{V} \in \mathbb{R}^{L \times D}$ using linear projection matrices $\mathbf{W_Q}, \mathbf{W_K}, \mathbf{W_V} \in \mathbb{R}^{D \times D}$ and they are input into LSA. The original $\mathbf{K}$ and $\mathbf{V}$ are projected into $\widetilde{\mathbf{K}}, \widetilde{\mathbf{V}} \in \mathbb{R}^{P \times D}$ through the item-to-interest aggregation function $f$:

$$\tilde{\mathbf{S}}_i = \text{softmax}\left(\frac{\mathbf{Q}_i \cdot \tilde{\mathbf{K}}_i^\top}{\sqrt{D/h}}\right)\tilde{\mathbf{V}}_i = \text{softmax}\left(\frac{\mathbf{Q}_i \cdot f(\mathbf{K}_i)^\top}{\sqrt{D/h}}\right) f(\mathbf{V}_i), \qquad (6)$$

where $h$ is the number of attention heads and $i$ is the index of the attention head. Multiple $\mathbf{S}_i$ are concatenated to obtain the final output of the LSA, $\tilde{\mathbf{S}} \in \mathbb{R}^{L \times D}$. $\tilde{\mathbf{S}}$ is used as the context-aware representation.

## 3 MLSA4Rec

In this section, we will first introduce the overall framework of MLSA4Rec. Then, we will gradually introduce the important components that constitute MLSA4Rec and demonstrate how to build a sequential recommendation model using these components. Finally, the complexity of MLSA4Rec is analyzed.

### 3.1 Framework Overview

As shown in Figure 1, our proposed MLSA4Rec is a novel hybrid sequential recommendation model that combines the advantages of the LSA and Mamba. MLSA4Rec consists of four main components: the embedding layer, the Mamba-LSA interaction layer, the Mamba normalization layer, and the prediction layer. The embedding layer receives the user's historical interaction sequence and transforms it into an embedding vector representation. The Mamba-LSA interaction layer is the core component of MLSA4Rec, which comprehensively captures user preference information through the synergy of the LSA module and Mamba block. The Mamba normalization layer further processes and refines the output of the Mamba-LSA interaction module. Finally, the prediction layer generates the prediction results for the user's next possible interaction. The following sections will provide detailed introductions to these components.

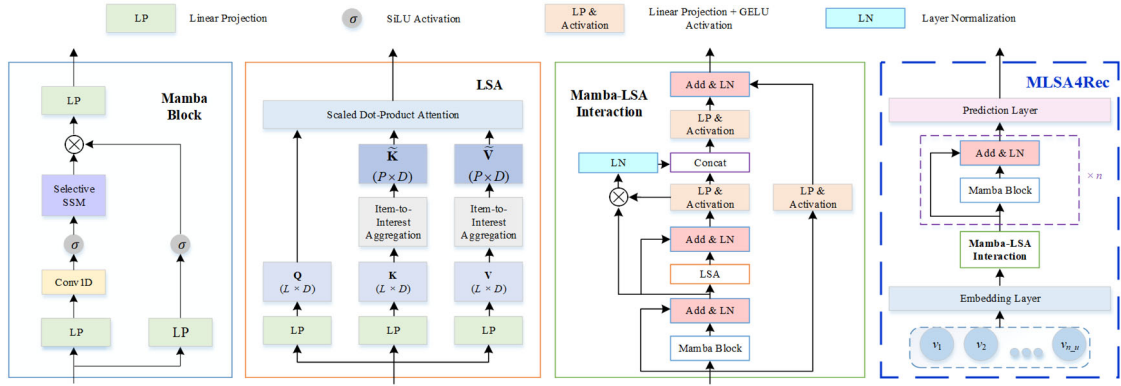

**Figure 1: The overview of MLSA4Rec.**

### 3.2 Embedding Layer

The embedding layer maps item IDs to the corresponding $D$-dimensional embedding space through a learnable embedding matrix $\mathbf{M} \in \mathbb{R}^{|V| \times D}$, where $|V|$ represents the number of items in the item set $V$. Similar to most existing works [11,16], the user's interaction sequence is normalized to a fixed length $L$ input sequence before being fed into the embedding layer. After passing through the embedding layer, we obtain the user's input embedding matrix $\mathbf{E} \in \mathbb{R}^{L \times D}$.

### 3.3 Mamba-LSA Interaction Layer

This layer is the core of MLSA4Rec. At the bottom of this layer is a Mamba block. After processing by this block, the Mamba block aggregates both global and sequential information into a whole, which serves as the input for the LSA module:

$$\mathbf{H} = \mathrm{LN}_1(\mathrm{Mamba}(\mathbf{E}) + \mathbf{E}) \tag{7}$$

where $\mathbf{H} \in \mathbb{R}^{L \times D}$, and $\mathrm{LN}_m(\cdot)$ represents the *m*-th layer normalization layer [27]. In other words, this corresponds to the structural bias injected into the LSA module by Mamba. The LSA leverages the low-rank property of the user interaction sequence for reducing the computational complexity of self-attention:

$$\mathbf{H}_{\mathrm{LSA}} = \mathrm{LN}_2(\mathrm{LSA}(\mathbf{H}) + \mathbf{H}) \tag{8}$$

Subsequently, LSA guides Mamba to focus on important information in the user's historical interactions through a gated information transmission mechanism, based on its understanding of the user's potential interests:

$$\mathbf{H}_M = \mathrm{LN}_3(\mathbf{H} \odot \mathrm{MLP}_1(\mathbf{H}_{\mathrm{LSA}})) = \mathrm{LN}_3(\mathbf{H} \odot \mathrm{GELU}(\mathrm{Linear}(\mathbf{H}_{\mathrm{LSA}}))) \tag{9}$$

where $\odot$ represents element-wise multiplication. This realizes the information exchange between Mamba and LSA.

After completing the interaction between Mamba and LSA, the Mamba-LSA interaction layer integrates $\mathbf{H}_{\mathrm{LSA}}$ and $\mathbf{H}_M \in \mathbb{R}^{L \times D}$, as well as the embedding matrix $\mathbf{E}$, through linear projections and GELU activation [28] functions:

$$\mathbf{H}_{\mathrm{MLSA}} = \mathrm{LN}_4(\mathrm{MLP}_2(\mathrm{Concat}(\mathbf{H}_M, \mathrm{MLP}_1(\mathbf{H}_{\mathrm{LSA}}))) + \mathrm{MLP}_3(\mathbf{E})) \tag{10}$$

where $\mathrm{Concat}(\cdot)$ represents the concatenation operation.

### 3.4 Mamba Normalization Layer

In the Mamba-LSA interaction layer, the LSA module and the Mamba block have already provided valuable information through their interaction. The Mamba normalization layer can further process, integrate, and refine this information, ensuring that information from different sources is fully utilized and can more accurately reflect the user's preferences. Specifically, we stack $n$ Mamba normalization layers. The definition of the *b*-th Mamba normalization layer is as follows:

$$\mathbf{H}^{(b)} = \mathrm{LN}^{(b)}(\mathrm{Mamba}^{(b)}(\mathbf{H}^{(b-1)}) + \mathbf{H}^{(b-1)}), \quad \forall b \in \{1, 2, \ldots, n\} \tag{11}$$

where $\mathbf{H}^{(0)} = \mathbf{H}_{\mathrm{MLSA}}$. After passing through $n$ Mamba normalization layers, the final output $\widehat{\mathbf{H}} = \mathbf{H}^{(n)} \in \mathbb{R}^{L \times D}$ is obtained.

### 3.5 Prediction Layer

In the prediction layer, we utilize the embedding $h_{last} = \widehat{\mathbf{H}}_L \in \mathbb{R}^D$ of the last interacted item from the Mamba normalization layer as the representation of the user's current interest. This representation is employed to predict the probability scores for items in the item set $V$ being interacted with:

$$y_{pred} = \mathrm{softmax}(\mathrm{Linear}(h_{last})) \tag{12}$$

where $y_{pred} \in \mathbb{R}^{|V|}$ denotes the probability distribution of the items in $V$ that may serve as the next interaction item.

Through the above architecture design, MLSA4Rec can effectively capture comprehensive and multi-perspective dependencies among user interaction items. This occurs in cases where complexity scales linearly with the length of user historical interaction sequences, thereby improving the performance of sequence recommendation.

### 3.6 Complexity Analysis

The complexity of the MLSA4Rec model scales linearly with the length of the sequence. Here is a detailed analysis:

(1) Low-Rank Decomposition Self-Attention (LSA) Module: This module reduces computational load by projecting items into $P$ latent interests, decreasing the complexity of self-attention from $O(L^2)$ to $O(PL)$. Since $P$ is typically much smaller than $L$, this transformation significantly reduces complexity, maintaining a linear relationship with sequence length $L$.

(2) Mamba Block: The Mamba block implements a selective state space model with linear computational complexity through efficient hardware-aware design.

(3) Mamba-LSA Interaction Layer: This layer, the core of MLSA4Rec, combines LSA and Mamba modules. Both modules maintain linear computational complexity individually, and other transformations within the Mamba-LSA interaction layer do not introduce quadratic complexity, thus preserving linear complexity with sequence length.

(4) Mamba Normalization Layer: This layer further refines information using the Mamba block and layer normalization, maintaining linear complexity with sequence length.

In summary, MLSA4Rec's computational complexity scales linearly with the length of user historical interaction sequences.

## 4 Experiments

### 4.1 Dataset

We conducted experiments on three real-world datasets: MovieLens-1M, Amazon-Beauty, and Amazon-Video-Games. These datasets cover diverse domains such as movies, beauty products, and video games, ensuring high representativeness. Consistent with prior work [16], we filtered users and items with fewer than 5 interactions from the datasets and adopted a leave-one-out training, validation, and testing set split. After preprocessing, detailed statistics of the datasets can be found in Table 1.

**Table 1: Dataset statistics.**

| Dataset | # Users | # Items | # Interactions | Avg. Length |
|---|---|---|---|---|
| MovieLens-1M | 6,040 | 3,416 | 999,611 | 165.4 |
| Amazon-Beauty | 22,363 | 12,101 | 198,502 | 8.9 |
| Amazon-Video-Games | 14,494 | 6,950 | 132,209 | 9.1 |

### 4.2 Experimental Setup

We compared MLSA4Rec with several methods: GRU4Rec [8], NARM [9], SASRec [11], BERT4Rec [12], and Mamba4Rec [16]. Evaluation metrics include HR@10, NDCG@10, and MRR@10. The learning rate for all datasets was set to 0.001, maximum sequence length $L$ was set to 50, and the dimensionality of hidden layers (channels) was set to 64. Regarding parameters for the Mamba block, following the work [16], the SSM state expansion factor was set to 32, the kernel size of one-dimensional convolution was set to 4, and the block expansion factor for linear projection was set to 2. Other hyperparameters such as batch size, number of Mamba normalization layers $n$, dropout rate, number of attention heads $h$, and number of latent interests $P$ were determined based on grid search results on each dataset. The datasets used by MLSA4Rec and their split are the same as the work [16], and the algorithmic implementations of both MLSA4Rec and Mamba4Rec are based on the popular open-source recommendation library RecBole [29]. So the experimental results for GRU4Rec, NARM, SASRec, BERT4Rec, and Mamba4Rec were taken directly from the work [16]. We conducted four independent experiments on each dataset for MLSA4Rec and averaged the results.

### 4.3 Overall Performance

The overall performance of MLSA4Rec with the comparison method is shown in Table 2. On the MovieLens-1M dataset, MLSA4Rec significantly outperforms the other compared methods, indicating that it can effectively capture users' sequential behavioral patterns better. On the Amazon-Beauty and Amazon-Video-Games datasets, although the improvement of MLSA4Rec is not as significant as on MovieLens-1M, it still outperforms other methods in most metrics. These results demonstrate the robustness of MLSA4Rec in handling sparse datasets and further confirm its broad applicability and advantages across diverse datasets.

### 4.4 Ablation Studies

To verify the necessity of each component in MLSA4Rec, we conducted ablation studies. Table 3 shows the performance of MLSA4Rec and its variants on the MovieLens-1M dataset. The following are the descriptions and performance analyses of each variant:

**Table 2: Comparison of Recommendation Performance (optimal and sub-optimal performance are bolded and underlined respectively).**

| Method | MovieLens-1M | | | Amazon-Beauty | | | Amazon-Video-Games | | |
| --- | --- | --- | --- | --- | --- | --- | --- | --- | --- |
| | HR | NDCG | MRR | HR | NDCG | MRR | HR | NDCG | MRR |
| NARM | 0.2735 | 0.1506 | 0.1132 | 0.0627 | 0.0347 | 0.0262 | 0.1032 | 0.0530 | 0.0379 |
| GRU4Rec | 0.2934 | 0.1642 | 0.1249 | 0.0606 | 0.0332 | 0.0249 | 0.1030 | 0.0536 | 0.0380 |
| SASRec | 0.2977 | 0.1687 | 0.1294 | **0.0847** | 0.0425 | 0.0296 | <u>0.1168</u> | 0.0571 | 0.0390 |
| BERT4Rec | 0.3098 | 0.1764 | 0.1357 | 0.0760 | 0.0393 | 0.0282 | 0.1053 | 0.0538 | 0.0381 |
| Mamba4Rec | <u>0.3121</u> | <u>0.1822</u> | <u>0.1425</u> | <u>0.0812</u> | <u>0.0451</u> | <u>0.0362</u> | 0.1152 | <u>0.0603</u> | <u>0.0438</u> |
| MLSA4Rec | **0.3263** | **0.1929** | **0.1520** | 0.0768 | **0.0465** | **0.0373** | **0.1183** | **0.0645** | **0.0483** |

- **V1 (Replacing Mamba-LSA Interaction Layer)**: Replaces the Mamba-LSA interaction layer with one Mamba normalization layer.
- **V2 (Removing Mamba Blocks)**: Removes all Mamba blocks from MLSA4Rec and replaces the Mamba normalization layers with the same number of point-wise feed-forward networks (PFFNs).
- **V3 (Replacing LSA)**: Replaces LSA in the Mamba-LSA interaction layer with the vanilla self-attention.

- **V4 (Replacing Mamba Normalization Layer)**: Replaces only the Mamba normalization layers with the same number of PFFNs.

Table 3: Ablation Analysis on MovieLens-1M.

| Architecture | MovieLens-1M | | |
|---|---|---|---|
| | HR@10 | NDCG@10 | MRR@10 |
| V1 | 0.3187 | 0.1837 | 0.1424 |
| V2 | 0.2437 | 0.1327 | 0.0989 |
| V3 | 0.3191 | 0.1903 | 0.151 |
| V4 | 0.3167 | 0.189 | 0.15 |
| Default | **0.3263** | **0.1929** | **0.152** |

The performance comparison of V1, V2, and the default MLSA4Rec architecture indicates that the Mamba and self-attention mechanisms are complementary, and neither standalone Mamba nor LSA can model users' sequential behavior patterns as effectively as their combination. Among all variants, V3 is the closest in performance to the default architecture, but LSA in MLSA4Rec slightly outperforms the vanilla self-attention mechanism. This may be because it avoids direct interactions between items, making the model more resistant to noise in the sequence and improving generalization performance. The comparison between V4 and the default architecture highlights the necessity of the Mamba normalization layer and its ability to further mine user preferences.

In summary, MLSA4Rec effectively enhances sequential recommendation performance by combining LSA and Mamba. Experimental results fully validate the model's broad applicability and superiority across different datasets.

## 5 Conclusion

In this paper, we propose a novel hybrid sequential recommendation framework called MLSA4Rec. By combining the Mamba model with a low-rank decomposition self-attention mechanism, MLSA4Rec efficiently captures comprehensive and multi-perspective dependencies within user interaction sequences. Experimental results have shown that our method outperforms many competitive baselines across multiple real-world datasets, showcasing the great potential of the synergistic operation between the Mamba and self-attention mechanisms. Future research can further optimize the Mamba-LSA interaction module and explore its applications in other recommendation tasks.

## REFERENCES

[1] Wang, Z., Zou, Y., Dai, A., Hou, L., Qiao, N., Zou, L., Ma, M., Ding, Z. and Xu, S., 2023, September. An Industrial Framework for Personalized Serendipitous Recommendation in E-commerce. In *Proceedings of the 17th ACM Conference on Recommender Systems* (pp. 1015-1018).

[2] Deng, W., Zhu, P., Chen, H., Yuan, T. and Wu, J., 2023. Knowledge-aware sequence modelling with deep learning for online course recommendation. *Information Processing & Management*, *60*(4), p.103377.

[3] Moscati, M., Wallmann, C., Reiter-Haas, M., Kowald, D., Lex, E. and Schedl, M., 2023, September.

Integrating the act-r framework with collaborative filtering for explainable sequential music recommendation. In *Proceedings of the 17th ACM conference on recommender systems* (pp. 840-847).

[4] Rendle, S., Freudenthaler, C. and Schmidt-Thieme, L., 2010, April. Factorizing personalized markov chains for next-basket recommendation. In *Proceedings of the 19th international conference on World wide web* (pp. 811-820).

[5] He, R. and McAuley, J., 2016, December. Fusing similarity models with markov chains for sparse sequential recommendation. In *2016 IEEE 16th international conference on data mining (ICDM)* (pp. 191-200). IEEE.

[6] Tang, J. and Wang, K., 2018, February. Personalized top-n sequential recommendation via convolutional sequence embedding. In *Proceedings of the eleventh ACM international conference on web search and data mining* (pp. 565-573).

[7] Yan, A., Cheng, S., Kang, W.C., Wan, M. and McAuley, J., 2019, November. CosRec: 2D convolutional neural networks for sequential recommendation. In *Proceedings of the 28th ACM international conference on information and knowledge management* (pp. 2173-2176).

[8] Hidasi, B., Karatzoglou, A., Baltrunas, L. and Tikk, D., 2015. Session-based recommendations with recurrent neural networks. *arXiv preprint arXiv:1511.06939*.

[9] Li, J., Ren, P., Chen, Z., Ren, Z., Lian, T. and Ma, J., 2017, November. Neural attentive session-based recommendation. In *Proceedings of the 2017 ACM on Conference on Information and Knowledge Management* (pp. 1419-1428).

[10] Kirkpatrick, J., Pascanu, R., Rabinowitz, N., Veness, J., Desjardins, G., Rusu, A.A., Milan, K., Quan, J., Ramalho, T., Grabska-Barwinska, A. and Hassabis, D., 2017. Overcoming catastrophic forgetting in neural networks. *Proceedings of the national academy of sciences*, *114*(13), pp.3521-3526.

[11] Kang, W.C. and McAuley, J., 2018, November. Self-attentive sequential recommendation. In *2018 IEEE international conference on data mining (ICDM)* (pp. 197-206). IEEE.

[12] Sun, F., Liu, J., Wu, J., Pei, C., Lin, X., Ou, W. and Jiang, P., 2019, November. BERT4Rec: Sequential recommendation with bidirectional encoder representations from transformer. In *Proceedings of the 28th ACM international conference on information and knowledge management* (pp. 1441-1450).

[13] Zuo, S., Liu, X., Jiao, J., Charles, D., Manavoglu, E., Zhao, T. and Gao, J., 2022. Efficient long sequence modeling via state space augmented transformer. *arXiv preprint arXiv:2212.08136*.

[14] Fan, X., Liu, Z., Lian, J., Zhao, W.X., Xie, X. and Wen, J.R., 2021, July. Lighter and better: low-

rank decomposed self-attention networks for next-item recommendation. In *Proceedings of the 44th international ACM SIGIR conference on research and development in information retrieval* (pp. 1733-1737).

[15] Lin, T., Wang, Y., Liu, X. and Qiu, X., 2022. A survey of transformers. *AI open*, *3*, pp.111-132.

[16] Liu, C., Lin, J., Wang, J., Liu, H. and Caverlee, J., 2024. Mamba4rec: Towards efficient sequential recommendation with selective state space models. *arXiv preprint arXiv:2403.03900*.

[17] Wang, Y., He, X. and Zhu, S., 2024. EchoMamba4Rec: Harmonizing Bidirectional State Space Models with Spectral Filtering for Advanced Sequential Recommendation. *arXiv preprint arXiv:2406.02638*.

[18] Gu, A. and Dao, T., 2023. Mamba: Linear-time sequence modeling with selective state spaces. *arXiv preprint arXiv:2312.00752*.

[19] Xu, X., Liang, Y., Huang, B., Lan, Z. and Shu, K., 2024. Integrating Mamba and Transformer for Long-Short Range Time Series Forecasting. *arXiv preprint arXiv:2404.14757*.

[20] Park, J., Park, J., Xiong, Z., Lee, N., Cho, J., Oymak, S., Lee, K. and Papailiopoulos, D., 2024. Can mamba learn how to learn? a comparative study on in-context learning tasks. *arXiv preprint arXiv:2402.04248*.

[21] Lieber, O., Lenz, B., Bata, H., Cohen, G., Osin, J., Dalmedigos, I., Safahi, E., Meirom, S., Belinkov, Y., Shalev-Shwartz, S. and Abend, O., 2024. Jamba: A hybrid transformer-mamba language model. *arXiv preprint arXiv:2403.19887*.

[22] Vaswani, A., Shazeer, N., Parmar, N., Uszkoreit, J., Jones, L., Gomez, A.N., Kaiser, Ł. and Polosukhin, I., 2017. Attention is all you need. *Advances in neural information processing systems*, *30*.

[23] Kalman, R.E., 1960. A new approach to linear filtering and prediction problems.

[24] Patro, B.N. and Agneeswaran, V.S., 2024. Mamba-360: Survey of state space models as transformer alternative for long sequence modelling: Methods, applications, and challenges. *arXiv preprint arXiv:2404.16112*.

[25] Gu, A., Goel, K. and Ré, C., 2021. Efficiently modeling long sequences with structured state spaces. *arXiv preprint arXiv:2111.00396*.

[26] Xu, R., Yang, S., Wang, Y., Du, B. and Chen, H., 2024. A survey on vision mamba: Models, applications and challenges. *arXiv preprint arXiv:2404.18861*.

[27] Ba, J.L., Kiros, J.R. and Hinton, G.E., 2016. Layer normalization. *arXiv preprint arXiv:1607.06450*.

[28] Hendrycks, D. and Gimpel, K., 2016. Gaussian error linear units (gelus). *arXiv preprint*

*arXiv:1606.08415*.

[29] Zhao, W.X., Mu, S., Hou, Y., Lin, Z., Chen, Y., Pan, X., Li, K., Lu, Y., Wang, H., Tian, C. and Min, Y., 2021, October. Recbole: Towards a unified, comprehensive and efficient framework for recommendation algorithms. In *proceedings of the 30th acm international conference on information & knowledge management* (pp. 4653-4664).